 \documentclass[final,5p,times,twocolumn]{elsarticle}
 \biboptions{numbers,sort&compress}

\usepackage{amssymb}
\usepackage{lipsum}

\usepackage{amsmath}
\usepackage{amssymb}
\usepackage{graphicx}
\usepackage{dcolumn}
\usepackage{bm}
\usepackage{braket}
\usepackage{mathrsfs}
\usepackage{booktabs}
\usepackage{multirow}
\usepackage{url}
\usepackage{physics}
\usepackage{subfigure}
\usepackage[colorlinks=true,linkcolor=blue, filecolor=blue,urlcolor=blue,citecolor=blue]{hyperref}
\usepackage{color}
\usepackage{cuted}
\usepackage{lipsum}

\journal{Physics Letters B}

\begin{document}

\begin{frontmatter}

\title{Data-driven trap theory for nuclear scattering}

\author[first]{Hantao Zhang}
\author[second,fourth]{Dong Bai}
\author[fifth]{Xilin Zhang}
\author[first,third]{Zhongzhou Ren\corref{cor1}}
\ead{zren@tongji.edu.cn}
\cortext[cor1]{Corresponding author}

\affiliation[first]{organization={School of Physics Science and Engineering},
            addressline={Tongji University}, 
            city={Shanghai},
            postcode={200092}, 
            country={China}}
\affiliation[second]{organization={College of Mechanics and Engineering Science},
    addressline={Hohai University}, 
    city={Nanjing},
    postcode={211100}, 
    country={China}}
\affiliation[third]{organization={Key Laboratory of Advanced Micro-Structure Materials},
    addressline={Tongji University}, 
    city={Shanghai},
    postcode={200092}, 
    country={China}}
\affiliation[fourth]{organization={Shanghai Research Center for Theoretical Nuclear Physics},
	addressline={NSFC and Fudan University}, 
	city={Shanghai},
	postcode={200438}, 
	country={China}}

\affiliation[fifth]{organization={Facility for Rare Isotope Beams, Michigan State University },
	postcode={MI 48824}, 
	country={USA}}
    
\begin{abstract}
We present a novel data-driven trap theory (abbreviated as DDTT) for nuclear scattering, which aims to overcome the limitations of the traditional trap method in dealing with narrow potential wells, while also providing a more efficient framework for handling long-range Coulomb interactions.
 As proof-of-concept examples, we employ this unified theory to analyze the elastic scattering of nucleon-nucleon  and nucleon-$\alpha$ systems. DDTT can successfully produce results consistent with those from traditional approaches, highlighting its significance for $\emph{ab initio}$ light nuclei scattering studies and  potential for applications in the heavier mass region.

\end{abstract}


\end{frontmatter}

\section{Introduction}
\label{introduction}

The trap method provides a powerful framework for extracting scattering phase shifts from the discrete energy spectra of confined systems, with broad applications across atomic physics, nuclear physics, and lattice quantum chromodynamics. Established quantization formulas—such as the  Busch-Englert-Rzażewski-Wilkens (BERW) formula \cite{Busch:1998cey} in the harmonic oscillator trap \cite{Luu:2010hw,Stetcu:2010xq,Schafer:2022hzo,Bagnarol:2023crb,Zhang:2019cai,Zhang:2020rhz,Guo:2021uig,Guo:2021qfu,Guo:2021lhz,Guo:2021hrf,Zhang:2024mot,Zhang:2024vch,Bagnarol:2024rhq,Zhang:2024vmz,Zhang:2024ykg}, L{\"u}scher’s formula \cite{Luscher:1990ux} in the periodic cubic box, and hard-sphere boundary conditions \cite{Borasoy:2007vy,Vento:2015yja,Rokash:2015hra}—connect  discrete energy levels of confined system to free-space scattering observables through a critical assumption: traps modify only asymptotic boundary conditions while leaving short-range interactions unaffected. This assumption breaks down for narrow confinements where finite-size effects introduce systematic errors, and faces fundamental limitations in charged-particle systems where the long-range Coulomb interaction remains incompletely addressed. 
{\color{black} Although theoretical efforts have explored Coulomb corrections \cite{Beane:2014qha,Davoudi:2018qpl,Stellin:2020gst,Christ:2021guf,Yu:2022nzm,Bubna:2024izx,Guo:2021qfu,Guo:2021lhz,Guo:2021hrf,Zhang:2024mot,Zhang:2024vch,Bagnarol:2024rhq,Rojik:2025jcv,Bubna:2025odg,Bubna:2025gsd}, a unified treatment of the potential well size dependence in quantization conditions remains elusive, with attempted numerical method proposed for harmonic trap  such as perturbation expansion approach \cite{Guo:2021qfu,Zhang:2024mot,Zhang:2024vch} and mesh method based Green’s function solver \cite{Bagnarol:2024rhq,Rojik:2025jcv} exhibiting numerical instabilities near singularities inherent to trap formalisms.}

To overcome these limitations, we introduce a novel data-driven trap theory (DDTT) that combines spectral expansions via generalized zeta functions, renormalization through momentum projection operators, and training on stochastic ensembles of near-zero-range potentials. This framework delivers a unified quantization condition valid for arbitrary trap geometries (including strong confinement), the full complex energy plane, and crucially, systems with long-range Coulomb interactions. Our approach provides the first systematic correction for finite-width effects in charged-particle traps, yielding phase shifts with significantly suppressed systematic errors. 
{\color{black}The derived quantization condition exhibits exceptional analytic behavior near singularities where conventional methods exhibit  pathological instabilities difficult to avoid, DDTT enables more robust treatment of near-pole states in the confinded quantum system.}

We demonstrate DDTT’s efficacy through high-precision calculations of nucleon-nucleon and $\alpha$-nucleon scattering phase shifts, establishing its superiority over conventional methods in handling both strong confinement  and Coulomb complexities. This work not only resolves long-standing challenges in trap formalisms but also provides a foundation for future ab initio studies of nuclear scattering and reaction via the trap method. 

The article is structured as follows: Section II develops the theoretical formalism of DDTT for neutral and charged particles. Section III presents numerical results and analysis. Section IV summarizes conclusions and broader implications.

\section{Theoretical Formalism}\label{Theoretical Formalism}


%
%
%
%
%
%

\begin{figure*}[!htbp] 
	\centering
	{\includegraphics[width=0.9\textwidth]{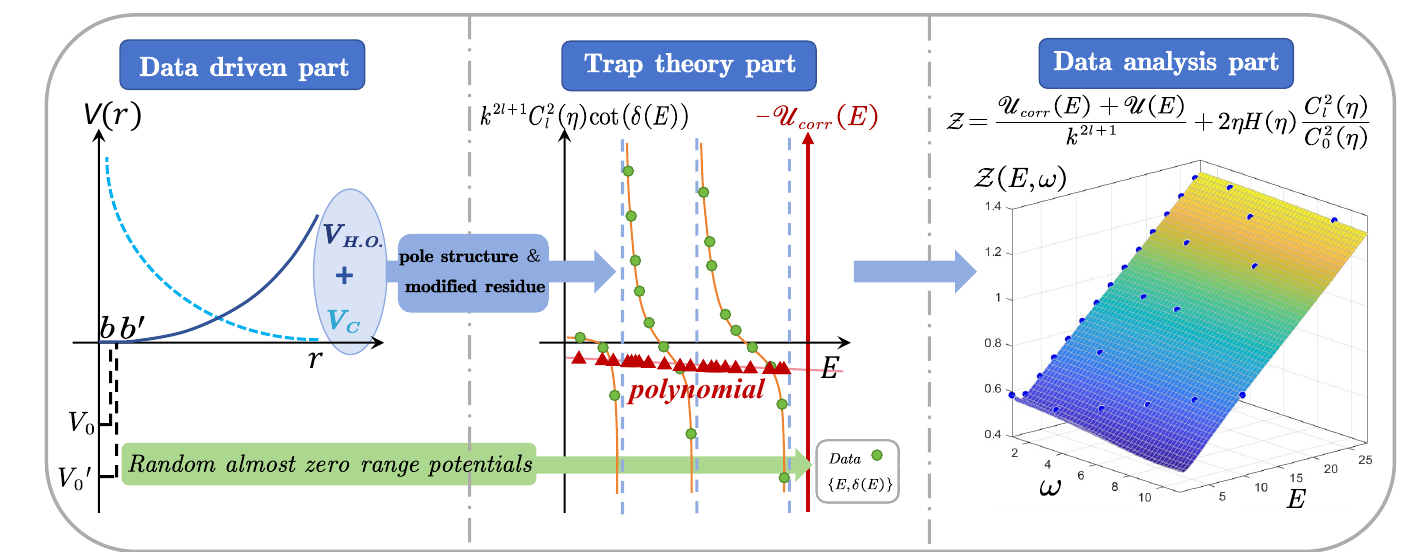} }
	\caption{
		Diagram of the data-driven trap theory with data analysis. As the frequency $\omega$ of the harmonic oscillator potential well increases, the size $\dfrac{\hbar}{\sqrt{\mu \omega}}$  of the trap decreases, which will lead to the breakdown of interaction independence. Consequently, a discrepancy arises between the Coulomb-corrected BERW formula and the exact phase shift. The data-driven trap theory with data analysis enables us to handle both large $\omega$ and charged particle scattering in a unified framework. In the data-driven part, the Coulomb and harmonic oscillator potentials provide the poles of trap formula and the modified residue with a certain momentum cutoff $\Lambda$. Then, using random almost zero range interactions we can obtain the data 
		$\{E_i,\delta(E_i)\}$. These data will be incorporated into the trap theory part to determine the correction function $U_{corr}$. Finally by introducing data analysis the contamination will be served as the modification to recalibrate the phase shifts of Coulomb-corrected BERW formula.  }	\label{DDTT} 
\end{figure*}

Generally, in the presence of Coulomb potential the pole structure of the finite volume $\zeta$ function leads to the following quantization condition in a artificial harmonic trap $\frac{\mu}{2\hbar^2}\omega^2r^2$, 
{\color{black}
\begin{equation}
\begin{aligned}
	&{C_{\ell}^2({\eta})} k^{2\ell+1} \cot\delta_{\ell}(E) = U_\omega(E)+U_{corr}(E) +\sum_{\alpha=0,\beta =1}^\infty b^S_{\alpha,\beta}E^\alpha 
	\omega^{2\beta}.
\end{aligned}
	\label{QC_trap_full}
\end{equation}
} 

On the left-hand side of Eq.(\ref{QC_trap_full}), $l$ is angular momentum, momentum $k$ is defined by $\frac{\sqrt{2\mu E}}{\hbar}$ with $\mu$ being two-body reduced mass, and $\delta_{\ell}$ is the phase shift in the free space. Coulomb wave function normalization factor $C^2_{\ell}(\eta)$ is defined as,
\begin{equation}\label{}
	\begin{aligned}
		&C^2_{\ell}(\eta)=C_0^2(\eta)\prod_{i=1}^{\ell}(1+\dfrac{\eta^2}{i^2}), \ \ \ C_0^2(\eta)=\dfrac{2\pi\eta}{e^{2\pi\eta}-1},
	\end{aligned}	
\end{equation}
where $\eta=\frac{Z_1Z_2e^2\mu}{\hbar^2k}$ is the Sommerfeld parameter, $Z_1$ and $Z_2$ are the charge numbers of two interacting particles. For charged particle scattering, the effective range expansion that will be associated with Eq.(\ref{QC_trap_full}) and data analysis can be expressed as,

\begin{equation}\label{}
	\begin{aligned}
 {C_{\ell}^2({\eta})} k^{2\ell+1} [\cot\delta_{\ell}(E)+\dfrac{2\eta H(\eta)}{C_0^2(\eta)}]= \sum_{\alpha=0}^\infty b^S_{\alpha} E^\alpha ,
	\end{aligned}	
\end{equation}
where function $H(\eta)$ is given by,
\begin{equation}
	\begin{aligned}
		H(\eta)&=\sum_{s=1}^{\infty}\frac{\eta^2}{s(s^2+\eta^2)}-\ln(\eta)-\gamma \\&=-\dfrac{i\pi}{e^{2\pi\eta}-1}+\Psi(i\eta)+\dfrac{1}{2i\eta}-\ln(i\eta) ,
	\end{aligned}
\end{equation}
$\gamma=0.5772156649\cdots$  is the Euler's constant, $\Psi(z)$ is the logarithmic derivative of the $\Gamma$ function ($\Psi$ function or digamma).

On the right-hand side of Eq.(\ref{QC_trap_full}), the first function $U_\omega$ is constructed by residue modified pole summation and momentum truncated Coulomb Green's function which serves as the renormalization term,
\begin{equation}
	\begin{aligned} 
		& U_{\omega}(E_a)=\dfrac{\hbar^2}{2\mu}\{\sum_i^{\infty} \frac{({R_i^{C}})^2}{E_a-E^{trap,C}_i} -
		\frac{2}{\pi} \mathscr{R} \int_0^\Lambda d p  p^2 \frac{\left[\dfrac{F_{\ell}(p,r)}{pr \times r^{\ell}}\right]_{r\to 0}^2}{ E - p^2/(2\mu) + i 0^+ }\}, 
	\end{aligned}
	\label{Uomegaterm}
\end{equation}
where ${R_i^{C}}$ is the modified residue which we will discuss later, $E^{trap,C}$ is the eigenenergy of the Hamiltonian $T+\frac{\mu}{2\hbar^2}\omega^2r^2+\frac{Z_1Z_2e^2}{r}$ and $\Lambda$ is the momentum cutoff. $\mathscr{R} $ denotes the real part of the integration, and the limitation $r\to 0$ can be strictly handled by using asymptotic behavior of Coulomb wave function.

The second function $U_{corr}$ is the correction term  determined by data-driven approach and the third summation correspond to  the  contamination term, which is introduced in the data-analysis \cite{Zhang:2019cai}  and served as modification term responsible for finite-size effects. $b^{S}$ in the last term denotes the coefficients dependent by the specific short range interaction $V_S$ and it should be noted that index $\beta$ is counted from 1 but not zero according to zero-$\omega$ limitation.

Returning to the discussion of modified residues, some analysis of the second term in Eq.(\ref{Uomegaterm}) reveals that the momentum cutoff in the Coulomb integration is essentially related to implementing a projection operator $\hat{P}$,
\begin{equation}
	\begin{aligned}
		\hat{P}_\Lambda \equiv\frac{2}{\pi} \int_0^{\Lambda} d p p^2 | p^+, \ell \rangle \langle p^+, \ell | ,
	\end{aligned}
\end{equation}
here $| p^+, \ell\rangle$ is the regular (outgoing) Coulomb wave function in the partial wave $\ell$.
With the defined  projection operator $\hat{P}_{\Lambda}$, the modified residue of charged case can be formulated for general partial waves as,

\begin{align}    
	R_i^{C}&=\dfrac{\langle r  ,\ell | \hat{P}_\Lambda | \phi_i \rangle }{r^{\ell}}|_{r\rightarrow 0}= \int_0^\Lambda \frac{2}{\pi} dp p^2 \dfrac{\langle r  , \ell | p^+,\ell \rangle}{r^{\ell}}|_{r\rightarrow 0} \langle p^+, \ell | \phi_i \rangle ,
\end{align}
where the limitation $r\to 0$ can be handled the same as the case in Eq.(\ref{Uomegaterm}). 





\begin{figure}[htbp] 
	\centering
	{\includegraphics[width=0.475
		\textwidth]{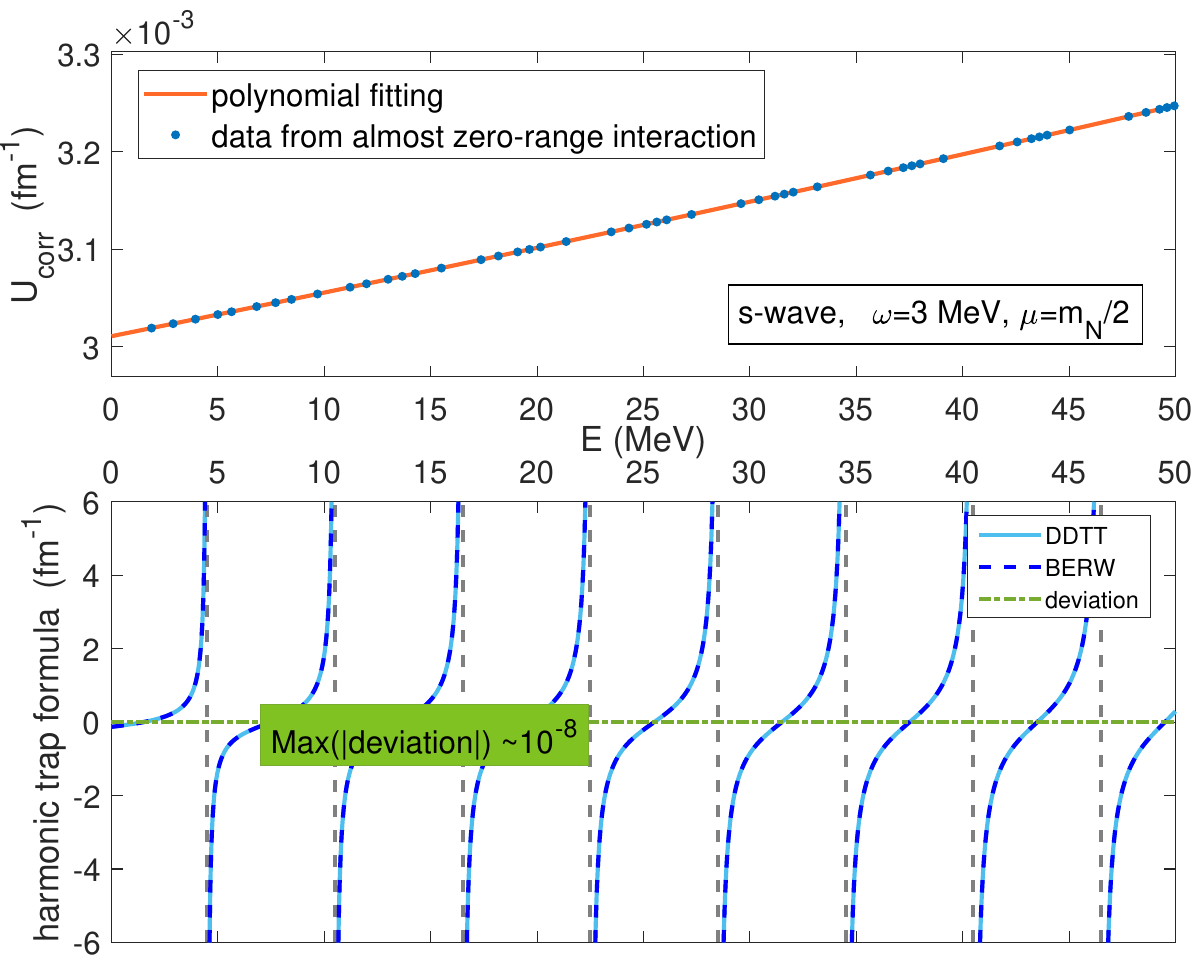} }	\caption{ Upper panel: The correction term $U_{corr}$ obtained via the data-driven approach. Data points are calculated using several almost zero-range interactions. The solid line represents the corresponding polynomial fitting.  Lower panel: Comparison between the harmonic trap formula derived from DDTT and the BERW formula. The solid line denotes the result from DDTT, while the dashed line represents the BERW result. The position of the divergence point is marked by  grey vertical dashed lines. The dot-dashed line indicates the difference between the two. Here, the orbital angular momentum is 0, the harmonic oscillator frequency is $\omega$ = 3 MeV, and the reduced mass is half the nucleon mass. The momentum cutoff $\Lambda=4.65$fm${}^{-1}$.}\label{data_driven_for_U_corr}
\end{figure}

In Fig. \ref{data_driven_for_U_corr} we display  the correction term $U_{corr}$ derived using the data-driven approach, along with a comparison to the BERW formula. It is evident that the trap formula derived using DDTT with a low-order polynomial fitting demonstrates pretty high accuracy.

{\color{black}It should be noted that although the modification process starts from some randomly selected almost zero-range interaction $V_S$, not every well depth of such short-range potential under the  Coulomb and harmonic trap fields can provide high-quality bound-state energy data. This is because when the bound-state energy lies too close to a divergence point, the resulting error amplification significantly undermines the reliability of polynomial fitting.}

{\color{black}Last but not least, the "data" introduced in our framework are not limited to those from almost short-range interaction potentials. Alternatively, one may directly compute the Green's function numerically to supply the necessary data (also away from divergence points). Regardless of the data source, the ultimate objective remains fine-tuning the trap formula based on pole summation, and consistent results can be achieved through either approach.}

\section{Numerical Results}\label{Numerical Results}

Although the quantum condition for neutral scattering is already known to be the closed-form BERW formula, as a preliminary  verification we first employ DDTT to reproduce the BERW formula through an alternative approach. For the neutral case, the pole position in Eq. (\ref{QC_trap_full}) reduces to the harmonic oscillator levels $2n+l+3/2$.
Derivations of both the residue and renormalization term can be easily completed.

\begin{figure}[htbp] 
	\centering
	{\includegraphics[width=0.45
		\textwidth]{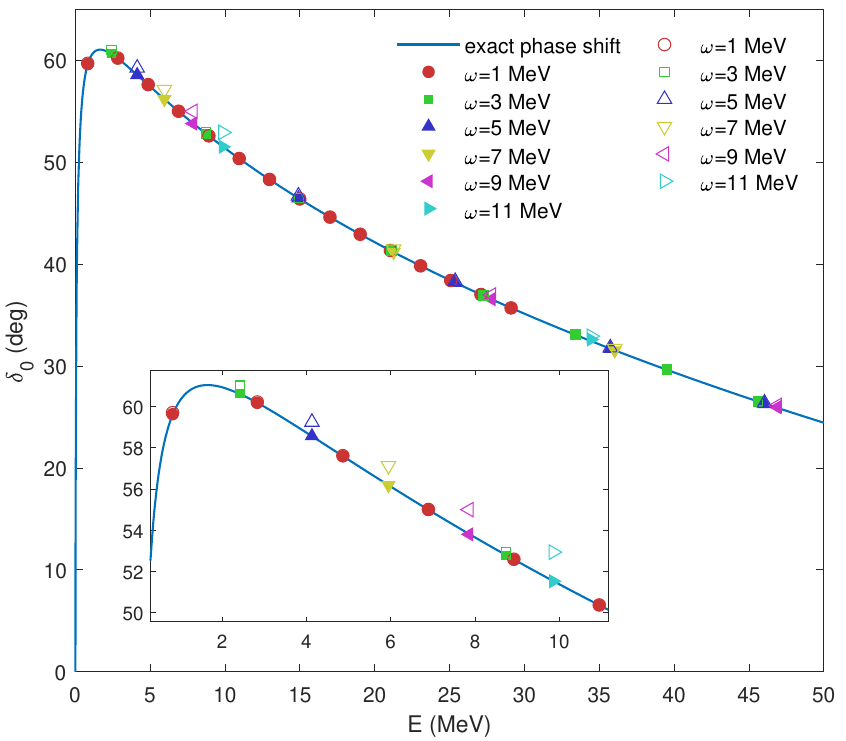} }	\caption{Neutron-neutron $s$-wave scattering phase shift with Av18 interaction. The solid curve represents the  results from $R$-matrix, while the hollow and solid circles correspond to the reproduced BERW and results obtained by DDTT, respectively.}\label{nnswave}
\end{figure}

As illustrated in Fig. \ref{DDTT}, the correction term $U_{corr}$ needs to be determined through  almost-zero-range interactions. In practical calculations, we select the smallest possible interaction range while maintaining numerical accuracy. For instance, in the following nucleon-nucleon scattering case, the chosen interaction range $b_0$ satisfies $\frac{b_0}{b_{\omega}}\approx 0.002 \ll 1$. Afterwards, $U_{corr}$ can be determined by randomly sampling multiple almost-zero-range interactions followed by low-order polynomial fitting.

{\color{black}In our calculations, we employed harmonic oscillator potentials with $\omega= 1, 3, 5, 7, 9, 11$ MeV for systematic data-analysis.}
{\color{black}It should also be noted that the quantization condition for the neutral case exhibits scale-invariant properties. Consequently, the reproduced BERW formula obtained for a specific reduced mass $\mu$ and harmonic oscillator frequency $\omega$ can be universally applied to any $\mu$ and $\omega$ through appropriate scaling - a feature that is also immediately evident from the closed-form BERW expression. However, for charged particle scattering, the presence of Coulomb interactions breaks this scale invariance. Therefore, the quantization conditions must be computed separately for each harmonic oscillator frequency $\omega$ within a given system (such as the proton-proton system discussed below).}

\begin{figure}[htbp] 
	\centering
	{\includegraphics[width=0.45
		\textwidth]{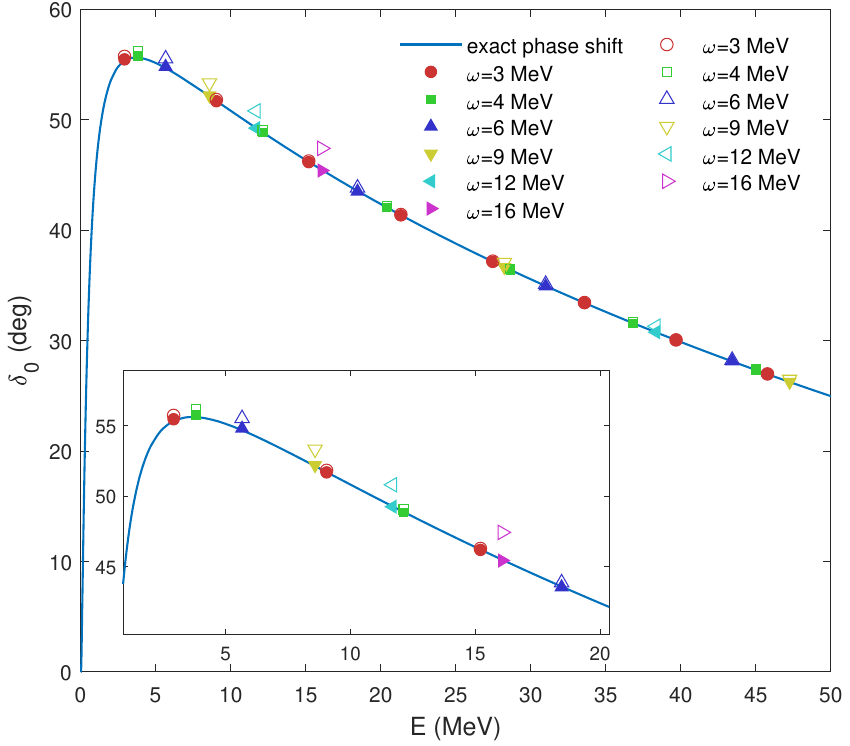} }	\caption{Proton-proton $s$-wave scattering phase shift with Av18 interaction. The solid curve represents the  results from $R$-matrix, while the  hollow and solid circles correspond to the Coulomb corrected-BERW and results obtained by DDTT, respectively.}\label{ppswave}
\end{figure}

Figure \ref{nnswave} presents the $s$-wave neutron-neutron scattering phase shifts obtained using DDTT with the Av18 interaction. The hollow markers represent results from the reproduced BERW formula, while the solid markers show the corrected results after data analysis. {\color{black}In the data-analysis fitting procedure, we expand the coefficients $b^S_{\alpha \beta}$ up to finite truncation.} The results in Fig.\ref{nnswave} demonstrate that DDTT effectively corrects for finite-size effects induced by potential well constraints.  Additionally, it should be noted that the discrepancy between the reproduced BERW and original BERW results is negligible, leading to nearly identical phase shifts that would overlap completely in the plot. For this reason, we have omitted the BERW data points from Fig. \ref{nnswave}.

After benchmarking with neutral particle scattering, we examine charged particle scattering using proton-proton scattering as a proof-of-concept example. 
Similar to the neutral scattering case, in Fig.\ref{ppswave} we display the $s$-wave proton-proton scattering phase shifts obtained using DDTT with the Av18 interaction. The hollow markers represent results from the Coulomb-corrected BERW formula, while the solid markers show the modified results after data analysis with harmonic oscillator frequency being $\omega= 3, 4,  6, 9, 12, 16$ MeV.  Our results demonstrate that DDTT approach successfully handles charged-particle scattering while effectively correcting for finite-size effects.

\begin{figure}[htbp] 
	\centering
	{\includegraphics[width=0.45
		\textwidth]{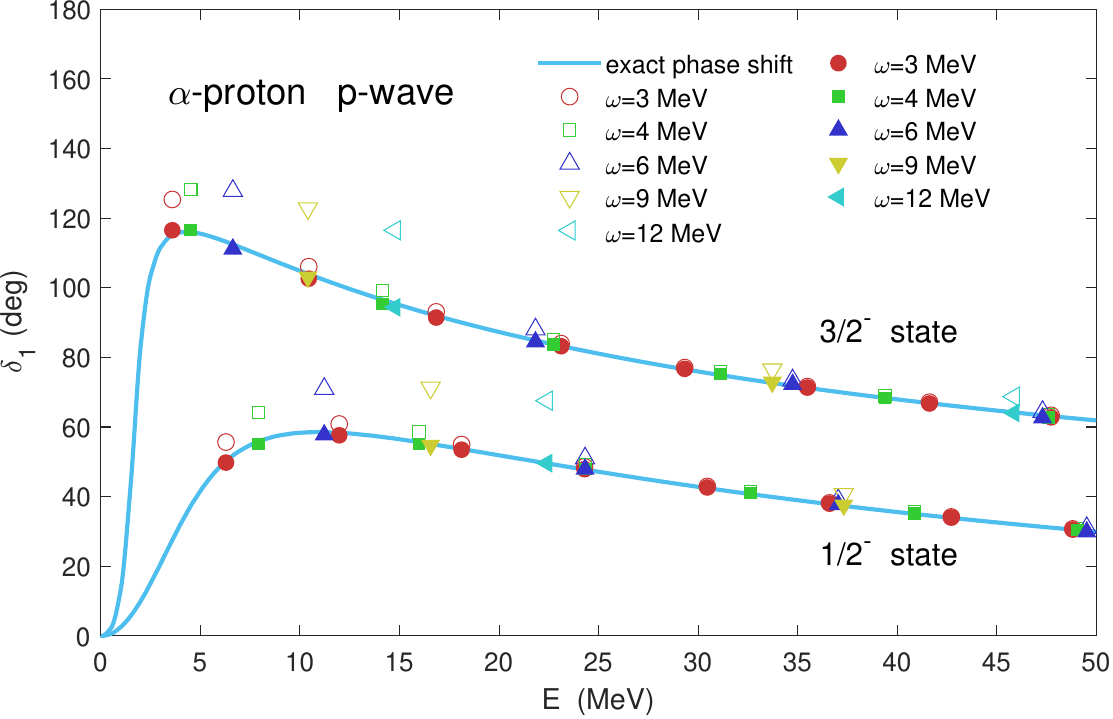} }	\caption{$\alpha$-proton phase shift scattering of $\frac{3}{2}^{-}$ and $\frac{1}{2}^{-}$ states. $\alpha-N$ effective interaction of Gaussian form is fitted from the results of microscopic cluster model. The solid curve represents the  results from $R$-matrix, while the open and solid circles correspond to the Coulomb corrected-BERW and results obtained by DDTT, respectively. }\label{palphapwave}
\end{figure}
{\color{black}Finally, for higher partial waves we take $p$-wave scattering  using the $\alpha$-proton system as a further examination.   Effective $\alpha$-nucleon interaction is obtained by fitting the results of microscopic cluster model. In Fig. \ref{palphapwave}, the hollow markers represent results from the Coulomb-corrected BERW formula, while the solid markers show the modified results after data analysis with harmonic oscillator frequency being $\omega= 3, 4,  6, 9, 12$ MeV. The results demonstrates that our DDTT approach maintains reliable phase shift estimation accuracy even for higher partial wave and system with larger charge magnitude.

In summary, DDTT delivers a unified theoretical framework for accurate phase-shift extraction, overcoming longstanding challenges in ab initio light nuclear studies via the trap method. Moreover, due to its favorable analytical properties, DDTT is expected to be further applied in handling complex interactions \cite{Zhang:2024ykg} and can also be utilized in quantum computing frameworks designed for non-Hermitian systems \cite{Zhang:2024rpa,Zhang:2025hsk,Zhang:2025hfm}.}

\section{Conclusions}\label{Conclusions}

{\color{black}
In this work we have introduced a novel data-driven trap theory (DDTT) based on the generalized Riemann $\zeta$ function, momentum dependent projection operator and data driven approach, which can provide a unified framework to effectively handle narrow potential trap, long-range Coulomb interactions, and various types of traps. By computing the scattering phase shifts for nucleon-nucleon and $\alpha$-nucleon systems, DDTT demonstrates strong consistency with traditional method. Its ability to successfully address strongly  confining traps and Coulomb potentials highlights the potential of DDTT in \textsl{ab initio} scattering research. Furthermore, this approach also shows promising applicability for nuclear scattering and resonance studies involving few-cluster system and heavier nuclei.}

\section*{Acknowledgements}
This work is supported by the National Natural Science Foundation of China (Grants No.\ 12535009, No.\ 124B2100, No.\ 11905103, No.\ 11947211, No.\ 11961141003, No.\ 12022517, No.\ 12375122 and No.\ 12147101), by the National Key R\&D Program of China (Contracts No. \ 2023YFA1606503)

\bibliographystyle{elsarticle-num-names} 
\bibliography{DDTT_reference}

\begin{thebibliography}{32}
\expandafter\ifx\csname natexlab\endcsname\relax\def\natexlab#1{#1}\fi
\providecommand{\url}[1]{\texttt{#1}}
\providecommand{\href}[2]{#2}
\providecommand{\path}[1]{#1}
\providecommand{\DOIprefix}{doi:}
\providecommand{\ArXivprefix}{arXiv:}
\providecommand{\URLprefix}{URL: }
\providecommand{\Pubmedprefix}{pmid:}
\providecommand{\doi}[1]{\href{http://dx.doi.org/#1}{\path{#1}}}
\providecommand{\Pubmed}[1]{\href{pmid:#1}{\path{#1}}}
\providecommand{\bibinfo}[2]{#2}
\ifx\xfnm\relax \def\xfnm[#1]{\unskip,\space#1}\fi
\bibitem[{Busch et~al.(1998)Busch, Englert, Rza{\.z}ewski, and
  Wilkens}]{Busch:1998cey}
\bibinfo{author}{T.~Busch}, \bibinfo{author}{B.-G. Englert},
  \bibinfo{author}{K.~Rza{\.z}ewski}, \bibinfo{author}{M.~Wilkens},
\newblock \bibinfo{title}{{Two Cold Atoms in a Harmonic Trap}},
\newblock \bibinfo{journal}{Found. Phys.} \bibinfo{volume}{28}
  (\bibinfo{year}{1998}) \bibinfo{pages}{549--559}.
  \DOIprefix\doi{10.1023/a:1018705520999}.
\bibitem[{Luu et~al.(2010)Luu, Savage, Schwenk, and Vary}]{Luu:2010hw}
\bibinfo{author}{T.~Luu}, \bibinfo{author}{M.~J. Savage},
  \bibinfo{author}{A.~Schwenk}, \bibinfo{author}{J.~P. Vary},
\newblock \bibinfo{title}{{Nucleon-Nucleon Scattering in a Harmonic
  Potential}},
\newblock \bibinfo{journal}{Phys. Rev. C} \bibinfo{volume}{82}
  (\bibinfo{year}{2010}) \bibinfo{pages}{034003}.
  \DOIprefix\doi{10.1103/PhysRevC.82.034003}.
  \href{http://arxiv.org/abs/1006.0427}{{\tt arXiv:1006.0427}}.
\bibitem[{Stetcu et~al.(2010)Stetcu, Rotureau, Barrett, and van
  Kolck}]{Stetcu:2010xq}
\bibinfo{author}{I.~Stetcu}, \bibinfo{author}{J.~Rotureau},
  \bibinfo{author}{B.~R. Barrett}, \bibinfo{author}{U.~van Kolck},
\newblock \bibinfo{title}{{An Effective field theory approach to two trapped
  particles}},
\newblock \bibinfo{journal}{Annals Phys.} \bibinfo{volume}{325}
  (\bibinfo{year}{2010}) \bibinfo{pages}{1644--1666}.
  \DOIprefix\doi{10.1016/j.aop.2010.02.008}.
  \href{http://arxiv.org/abs/1001.5071}{{\tt arXiv:1001.5071}}.
\bibitem[{Sch{\"a}fer and Bazak(2023)}]{Schafer:2022hzo}
\bibinfo{author}{M.~Sch{\"a}fer}, \bibinfo{author}{B.~Bazak},
\newblock \bibinfo{title}{{Few-nucleon scattering in pionless effective field
  theory}},
\newblock \bibinfo{journal}{Phys. Rev. C} \bibinfo{volume}{107}
  (\bibinfo{year}{2023}) \bibinfo{pages}{064001}.
  \DOIprefix\doi{10.1103/PhysRevC.107.064001}.
  \href{http://arxiv.org/abs/2208.10960}{{\tt arXiv:2208.10960}}.
\bibitem[{Bagnarol et~al.(2023)Bagnarol, Sch{\"a}fer, Bazak, and
  Barnea}]{Bagnarol:2023crb}
\bibinfo{author}{M.~Bagnarol}, \bibinfo{author}{M.~Sch{\"a}fer},
  \bibinfo{author}{B.~Bazak}, \bibinfo{author}{N.~Barnea},
\newblock \bibinfo{title}{{Five-body calculation of s-wave n-4He scattering at
  next-to-leading order pionless effective field theory}},
\newblock \bibinfo{journal}{Phys. Lett. B} \bibinfo{volume}{844}
  (\bibinfo{year}{2023}) \bibinfo{pages}{138078}.
  \DOIprefix\doi{10.1016/j.physletb.2023.138078}.
  \href{http://arxiv.org/abs/2306.04036}{{\tt arXiv:2306.04036}}.
\bibitem[{Zhang(2020)}]{Zhang:2019cai}
\bibinfo{author}{X.~Zhang},
\newblock \bibinfo{title}{{Extracting free-space observables from trapped
  interacting clusters}},
\newblock \bibinfo{journal}{Phys. Rev. C} \bibinfo{volume}{101}
  (\bibinfo{year}{2020}) \bibinfo{pages}{051602}.
  \DOIprefix\doi{10.1103/PhysRevC.101.051602}.
  \href{http://arxiv.org/abs/1905.05275}{{\tt arXiv:1905.05275}}.
\bibitem[{Zhang et~al.(2020)Zhang, Stroberg, Navr{\'a}til, Gwak, Melendez,
  Furnstahl, and Holt}]{Zhang:2020rhz}
\bibinfo{author}{X.~Zhang}, \bibinfo{author}{S.~R. Stroberg},
  \bibinfo{author}{P.~Navr{\'a}til}, \bibinfo{author}{C.~Gwak},
  \bibinfo{author}{J.~A. Melendez}, \bibinfo{author}{R.~J. Furnstahl},
  \bibinfo{author}{J.~D. Holt},
\newblock \bibinfo{title}{{Ab Initio Calculations of Low-Energy Nuclear
  Scattering Using Confining Potential Traps}},
\newblock \bibinfo{journal}{Phys. Rev. Lett.} \bibinfo{volume}{125}
  (\bibinfo{year}{2020}) \bibinfo{pages}{112503}.
  \DOIprefix\doi{10.1103/PhysRevLett.125.112503}.
  \href{http://arxiv.org/abs/2004.13575}{{\tt arXiv:2004.13575}}.
\bibitem[{Guo and Long(2022)}]{Guo:2021uig}
\bibinfo{author}{P.~Guo}, \bibinfo{author}{B.~Long},
\newblock \bibinfo{title}{{Nuclear reactions in artificial traps}},
\newblock \bibinfo{journal}{J. Phys. G} \bibinfo{volume}{49}
  (\bibinfo{year}{2022}) \bibinfo{pages}{055104}.
  \DOIprefix\doi{10.1088/1361-6471/ac59d5}.
  \href{http://arxiv.org/abs/2101.03901}{{\tt arXiv:2101.03901}}.
\bibitem[{Guo(2021)}]{Guo:2021qfu}
\bibinfo{author}{P.~Guo},
\newblock \bibinfo{title}{{Coulomb corrections to two-particle interactions in
  artificial traps}},
\newblock \bibinfo{journal}{Phys. Rev. C} \bibinfo{volume}{103}
  (\bibinfo{year}{2021}) \bibinfo{pages}{064611}.
  \DOIprefix\doi{10.1103/PhysRevC.103.064611}.
  \href{http://arxiv.org/abs/2101.11097}{{\tt arXiv:2101.11097}},
  \bibinfo{note}{[Erratum: Phys.Rev.C 111, 069903 (2025)]}.
\bibitem[{Guo and Gasparian(2021)}]{Guo:2021lhz}
\bibinfo{author}{P.~Guo}, \bibinfo{author}{V.~Gasparian},
\newblock \bibinfo{title}{{Charged particles interaction in both a finite
  volume and a uniform magnetic field}},
\newblock \bibinfo{journal}{Phys. Rev. D} \bibinfo{volume}{103}
  (\bibinfo{year}{2021}) \bibinfo{pages}{094520}.
  \DOIprefix\doi{10.1103/PhysRevD.103.094520}.
  \href{http://arxiv.org/abs/2101.01150}{{\tt arXiv:2101.01150}}.
\bibitem[{Guo and Gasparian(2022)}]{Guo:2021hrf}
\bibinfo{author}{P.~Guo}, \bibinfo{author}{V.~Gasparian},
\newblock \bibinfo{title}{{Charged particles interaction in both a finite
  volume and a uniform magnetic field II: topological and analytic properties
  of a magnetic system}},
\newblock \bibinfo{journal}{J. Phys. A} \bibinfo{volume}{55}
  (\bibinfo{year}{2022}) \bibinfo{pages}{265201}.
  \DOIprefix\doi{10.1088/1751-8121/ac7180}.
  \href{http://arxiv.org/abs/2107.10642}{{\tt arXiv:2107.10642}}.
\bibitem[{Zhang et~al.(2024{\natexlab{a}})Zhang, Bai, Wang, and
  Ren}]{Zhang:2024mot}
\bibinfo{author}{H.~Zhang}, \bibinfo{author}{D.~Bai},
  \bibinfo{author}{Z.~Wang}, \bibinfo{author}{Z.~Ren},
\newblock \bibinfo{title}{{Charged particle scattering in harmonic traps}},
\newblock \bibinfo{journal}{Phys. Lett. B} \bibinfo{volume}{850}
  (\bibinfo{year}{2024}{\natexlab{a}}) \bibinfo{pages}{138490}.
  \DOIprefix\doi{10.1016/j.physletb.2024.138490}.
\bibitem[{Zhang et~al.(2024{\natexlab{b}})Zhang, Bai, and Ren}]{Zhang:2024vch}
\bibinfo{author}{H.~Zhang}, \bibinfo{author}{D.~Bai}, \bibinfo{author}{Z.~Ren},
\newblock \bibinfo{title}{{Coupled-channels reactions for charged particles in
  harmonic traps}},
\newblock \bibinfo{journal}{Phys. Rev. C} \bibinfo{volume}{110}
  (\bibinfo{year}{2024}{\natexlab{b}}) \bibinfo{pages}{034308}.
  \DOIprefix\doi{10.1103/PhysRevC.110.034308}.
\bibitem[{Bagnarol et~al.(2025)Bagnarol, Barnea, Rojik, and
  Schafer}]{Bagnarol:2024rhq}
\bibinfo{author}{M.~Bagnarol}, \bibinfo{author}{N.~Barnea},
  \bibinfo{author}{M.~Rojik}, \bibinfo{author}{M.~Schafer},
\newblock \bibinfo{title}{{Accurate calculation of low energy scattering phase
  shifts of charged particles in a harmonic oscillator trap}},
\newblock \bibinfo{journal}{Phys. Lett. B} \bibinfo{volume}{861}
  (\bibinfo{year}{2025}) \bibinfo{pages}{139230}.
  \DOIprefix\doi{10.1016/j.physletb.2024.139230}.
  \href{http://arxiv.org/abs/2410.02602}{{\tt arXiv:2410.02602}}.
\bibitem[{Zhang et~al.(2024{\natexlab{a}})Zhang, Bai, Wang, and
  Ren}]{Zhang:2024vmz}
\bibinfo{author}{H.~Zhang}, \bibinfo{author}{D.~Bai},
  \bibinfo{author}{Z.~Wang}, \bibinfo{author}{Z.~Ren},
\newblock \bibinfo{title}{{Microscopic cluster model in harmonic oscillator
  traps}},
\newblock \bibinfo{journal}{Phys. Rev. C} \bibinfo{volume}{109}
  (\bibinfo{year}{2024}{\natexlab{a}}) \bibinfo{pages}{034307}.
  \DOIprefix\doi{10.1103/PhysRevC.109.034307}.
\bibitem[{Zhang et~al.(2024{\natexlab{b}})Zhang, Bai, and Ren}]{Zhang:2024ykg}
\bibinfo{author}{H.~Zhang}, \bibinfo{author}{D.~Bai}, \bibinfo{author}{Z.~Ren},
\newblock \bibinfo{title}{{Harmonic trap method for complex short-range
  potentials}},
\newblock \bibinfo{journal}{Phys. Lett. B} \bibinfo{volume}{855}
  (\bibinfo{year}{2024}{\natexlab{b}}) \bibinfo{pages}{138861}.
  \DOIprefix\doi{10.1016/j.physletb.2024.138861}.
\bibitem[{L{\"u}scher(1991)}]{Luscher:1990ux}
\bibinfo{author}{M.~L{\"u}scher},
\newblock \bibinfo{title}{{Two particle states on a torus and their relation to
  the scattering matrix}},
\newblock \bibinfo{journal}{Nucl. Phys. B} \bibinfo{volume}{354}
  (\bibinfo{year}{1991}) \bibinfo{pages}{531--578}.
  \DOIprefix\doi{10.1016/0550-3213(91)90366-6}.
\bibitem[{Borasoy et~al.(2007)Borasoy, Epelbaum, Krebs, Lee, and
  Meissner}]{Borasoy:2007vy}
\bibinfo{author}{B.~Borasoy}, \bibinfo{author}{E.~Epelbaum},
  \bibinfo{author}{H.~Krebs}, \bibinfo{author}{D.~Lee}, \bibinfo{author}{U.-G.
  Meissner},
\newblock \bibinfo{title}{{Two-particle scattering on the lattice: Phase
  shifts, spin-orbit coupling, and mixing angles}},
\newblock \bibinfo{journal}{Eur. Phys. J. A} \bibinfo{volume}{34}
  (\bibinfo{year}{2007}) \bibinfo{pages}{185--196}.
  \DOIprefix\doi{10.1140/epja/i2007-10500-9}.
  \href{http://arxiv.org/abs/0708.1780}{{\tt arXiv:0708.1780}}.
\bibitem[{Vento(2016)}]{Vento:2015yja}
\bibinfo{author}{V.~Vento},
\newblock \bibinfo{title}{{Glueball-Meson Mixing}},
\newblock \bibinfo{journal}{Eur. Phys. J. A} \bibinfo{volume}{52}
  (\bibinfo{year}{2016}) \bibinfo{pages}{1}.
  \DOIprefix\doi{10.1140/epja/i2016-16001-x}.
  \href{http://arxiv.org/abs/1505.05355}{{\tt arXiv:1505.05355}}.
\bibitem[{Rokash et~al.(2015)Rokash, Pine, Elhatisari, Lee, Epelbaum, and
  Krebs}]{Rokash:2015hra}
\bibinfo{author}{A.~Rokash}, \bibinfo{author}{M.~Pine},
  \bibinfo{author}{S.~Elhatisari}, \bibinfo{author}{D.~Lee},
  \bibinfo{author}{E.~Epelbaum}, \bibinfo{author}{H.~Krebs},
\newblock \bibinfo{title}{{Scattering cluster wave functions on the lattice
  using the adiabatic projection method}},
\newblock \bibinfo{journal}{Phys. Rev. C} \bibinfo{volume}{92}
  (\bibinfo{year}{2015}) \bibinfo{pages}{054612}.
  \DOIprefix\doi{10.1103/PhysRevC.92.054612}.
  \href{http://arxiv.org/abs/1505.02967}{{\tt arXiv:1505.02967}}.
\bibitem[{Beane and Savage(2014)}]{Beane:2014qha}
\bibinfo{author}{S.~R. Beane}, \bibinfo{author}{M.~J. Savage},
\newblock \bibinfo{title}{{Two-Particle Elastic Scattering in a Finite Volume
  Including QED}},
\newblock \bibinfo{journal}{Phys. Rev. D} \bibinfo{volume}{90}
  (\bibinfo{year}{2014}) \bibinfo{pages}{074511}.
  \DOIprefix\doi{10.1103/PhysRevD.90.074511}.
  \href{http://arxiv.org/abs/1407.4846}{{\tt arXiv:1407.4846}}.
\bibitem[{Davoudi et~al.(2019)Davoudi, Harrison, J{\"u}ttner, Portelli, and
  Savage}]{Davoudi:2018qpl}
\bibinfo{author}{Z.~Davoudi}, \bibinfo{author}{J.~Harrison},
  \bibinfo{author}{A.~J{\"u}ttner}, \bibinfo{author}{A.~Portelli},
  \bibinfo{author}{M.~J. Savage},
\newblock \bibinfo{title}{{Theoretical aspects of quantum electrodynamics in a
  finite volume with periodic boundary conditions}},
\newblock \bibinfo{journal}{Phys. Rev. D} \bibinfo{volume}{99}
  (\bibinfo{year}{2019}) \bibinfo{pages}{034510}.
  \DOIprefix\doi{10.1103/PhysRevD.99.034510}.
  \href{http://arxiv.org/abs/1810.05923}{{\tt arXiv:1810.05923}}.
\bibitem[{Stellin and Mei{\ss}ner(2021)}]{Stellin:2020gst}
\bibinfo{author}{G.~Stellin}, \bibinfo{author}{U.-G. Mei{\ss}ner},
\newblock \bibinfo{title}{{P-Wave Two-Particle Bound and Scattering States in a
  Finite Volume including QED}},
\newblock \bibinfo{journal}{Eur. Phys. J. A} \bibinfo{volume}{57}
  (\bibinfo{year}{2021}) \bibinfo{pages}{26}.
  \DOIprefix\doi{10.1140/epja/s10050-020-00319-1}.
  \href{http://arxiv.org/abs/2008.06553}{{\tt arXiv:2008.06553}}.
\bibitem[{Christ et~al.(2022)Christ, Feng, Karpie, and Nguyen}]{Christ:2021guf}
\bibinfo{author}{N.~Christ}, \bibinfo{author}{X.~Feng},
  \bibinfo{author}{J.~Karpie}, \bibinfo{author}{T.~Nguyen},
\newblock \bibinfo{title}{{{\ensuremath{\pi}}-{\ensuremath{\pi}} scattering,
  QED, and finite-volume quantization}},
\newblock \bibinfo{journal}{Phys. Rev. D} \bibinfo{volume}{106}
  (\bibinfo{year}{2022}) \bibinfo{pages}{014508}.
  \DOIprefix\doi{10.1103/PhysRevD.106.014508}.
  \href{http://arxiv.org/abs/2111.04668}{{\tt arXiv:2111.04668}}.
\bibitem[{Yu et~al.(2023)Yu, K{\"o}nig, and Lee}]{Yu:2022nzm}
\bibinfo{author}{H.~Yu}, \bibinfo{author}{S.~K{\"o}nig},
  \bibinfo{author}{D.~Lee},
\newblock \bibinfo{title}{{Charged-Particle Bound States in Periodic Boxes}},
\newblock \bibinfo{journal}{Phys. Rev. Lett.} \bibinfo{volume}{131}
  (\bibinfo{year}{2023}) \bibinfo{pages}{212502}.
  \DOIprefix\doi{10.1103/PhysRevLett.131.212502}.
  \href{http://arxiv.org/abs/2212.14379}{{\tt arXiv:2212.14379}}.
\bibitem[{Bubna et~al.(2024)Bubna, Hammer, M{\"u}ller, Pang, Rusetsky, and
  Wu}]{Bubna:2024izx}
\bibinfo{author}{R.~Bubna}, \bibinfo{author}{H.-W. Hammer},
  \bibinfo{author}{F.~M{\"u}ller}, \bibinfo{author}{J.-Y. Pang},
  \bibinfo{author}{A.~Rusetsky}, \bibinfo{author}{J.-J. Wu},
\newblock \bibinfo{title}{{L{\"u}scher equation with long-range forces}},
\newblock \bibinfo{journal}{JHEP} \bibinfo{volume}{05} (\bibinfo{year}{2024})
  \bibinfo{pages}{168}. \DOIprefix\doi{10.1007/JHEP05(2024)168}.
  \href{http://arxiv.org/abs/2402.12985}{{\tt arXiv:2402.12985}}.
\bibitem[{Rojik et~al.(2025)Rojik, Sch{\"a}fer, Bagnarol, and
  Barnea}]{Rojik:2025jcv}
\bibinfo{author}{M.~Rojik}, \bibinfo{author}{M.~Sch{\"a}fer},
  \bibinfo{author}{M.~Bagnarol}, \bibinfo{author}{N.~Barnea},
\newblock \bibinfo{title}{{Charged Particle Scattering in Renormalizable
  Pionless Effective Field Theory at Next-to-Leading Order: The $pd$, $dd$, and
  $p^3\mathrm{He}$ Case}}  (\bibinfo{year}{2025}).
  \href{http://arxiv.org/abs/2507.16250}{{\tt arXiv:2507.16250}}.
\bibitem[{Bubna et~al.(2025{\natexlab{a}})Bubna, Hammer, M{\"u}ller, Pang,
  Rusetsky, and Wu}]{Bubna:2025odg}
\bibinfo{author}{R.~Bubna}, \bibinfo{author}{H.-W. Hammer},
  \bibinfo{author}{F.~M{\"u}ller}, \bibinfo{author}{J.-Y. Pang},
  \bibinfo{author}{A.~Rusetsky}, \bibinfo{author}{J.-J. Wu},
\newblock \bibinfo{title}{{The finite-volume spectrum in the presence of a
  long-range force}},
\newblock \bibinfo{journal}{PoS} \bibinfo{volume}{LATTICE2024}
  (\bibinfo{year}{2025}{\natexlab{a}}) \bibinfo{pages}{092}.
  \DOIprefix\doi{10.22323/1.466.0092}.
\bibitem[{Bubna et~al.(2025{\natexlab{b}})Bubna, Hammer, Hoid, Pang, Rusetsky,
  and Wu}]{Bubna:2025gsd}
\bibinfo{author}{R.~Bubna}, \bibinfo{author}{H.-W. Hammer},
  \bibinfo{author}{B.-L. Hoid}, \bibinfo{author}{J.-Y. Pang},
  \bibinfo{author}{A.~Rusetsky}, \bibinfo{author}{J.-J. Wu},
\newblock \bibinfo{title}{{Modified L{\"u}scher zeta-function and the modified
  effective range expansion in the presence of a long-range force}}
  (\bibinfo{year}{2025}{\natexlab{b}}).
  \href{http://arxiv.org/abs/2507.18399}{{\tt arXiv:2507.18399}}.
\bibitem[{Zhang et~al.(2025{\natexlab{a}})Zhang, Bai, and Ren}]{Zhang:2024rpa}
\bibinfo{author}{H.~Zhang}, \bibinfo{author}{D.~Bai}, \bibinfo{author}{Z.~Ren},
\newblock \bibinfo{title}{{Quantum computing for extracting nuclear
  resonances}},
\newblock \bibinfo{journal}{Phys. Lett. B} \bibinfo{volume}{860}
  (\bibinfo{year}{2025}{\natexlab{a}}) \bibinfo{pages}{139187}.
  \DOIprefix\doi{10.1016/j.physletb.2024.139187}.
  \href{http://arxiv.org/abs/2409.06340}{{\tt arXiv:2409.06340}}.
\bibitem[{Zhang et~al.(2025{\natexlab{b}})Zhang, Bai, and Ren}]{Zhang:2025hsk}
\bibinfo{author}{H.~Zhang}, \bibinfo{author}{D.~Bai}, \bibinfo{author}{Z.~Ren},
\newblock \bibinfo{title}{{Iterative Harrow-Hassidim-Lloyd quantum algorithm
  for solving resonances with eigenvector continuation}}
  (\bibinfo{year}{2025}{\natexlab{b}}).
  \href{http://arxiv.org/abs/2506.20929}{{\tt arXiv:2506.20929}}.
\bibitem[{Zhang et~al.(2025{\natexlab{c}})Zhang, Bai, and Ren}]{Zhang:2025hfm}
\bibinfo{author}{H.~Zhang}, \bibinfo{author}{D.~Bai}, \bibinfo{author}{Z.~Ren},
\newblock \bibinfo{title}{{Studying few cluster resonances with quantum neural
  network driven iterative Harrow-Hassidim-Lloyd algorithm}}
  (\bibinfo{year}{2025}{\natexlab{c}}).
  \href{http://arxiv.org/abs/2507.00074}{{\tt arXiv:2507.00074}}.

\end{thebibliography}

\end{document}